\documentclass[twocolumn,nobibnotes,showpacs,superscriptaddress,aps,prl,longbibliography]{revtex4-2}

\usepackage{amsmath,graphicx,amssymb}
\usepackage[colorlinks,linkcolor=blue,urlcolor=blue,anchorcolor=blue,citecolor=blue]{hyperref}

\begin{document}

\title{Second order nonlinearity induced multipartite entanglement in a hybrid magnon cavity QED system}

\author{Y. Zhou}
\affiliation{MOE Key Laboratory of Advanced Micro-Structured Materials, School of Physics Science and Engineering, Tongji University, Shanghai 200092, China}
\affiliation{School of Electronics and Information Engineering, Taizhou University, Taizhou 318000, China}
\author{S. Y. Xie}
\email[Corresponding author:]{xieshuangyuan@tongji.edu.cn}
\affiliation{MOE Key Laboratory of Advanced Micro-Structured Materials, School of Physics Science and Engineering, Tongji University, Shanghai 200092, China}
\author{C. J. Zhu}
\email[Corresponding author:]{cjzhu@suda.edu.cn}
\affiliation{School of Physical Science and Technology, Soochow University, Suzhou 215006, China}
\affiliation{MOE Key Laboratory of Advanced Micro-Structured Materials, School of Physics Science and Engineering, Tongji University, Shanghai 200092, China}
\author{Y. P. Yang}
\email[Corresponding author:]{yaping\_yang@tongji.edu.cn}
\affiliation{MOE Key Laboratory of Advanced Micro-Structured Materials, School of Physics Science and Engineering, Tongji University, Shanghai 200092, China}

\begin{abstract}
We present a proposal to produce bipartite and tripartite entanglement in a hybrid magnon-cavity QED system. Two macroscopic yttrium iron garnet (YIG) spheres are coupled to a single-mode microwave cavity via magnetic dipole interaction, while the cavity photons are generated via the two photon process induced by a pump field. Using the mean field theory, we show that the second order nonlinearity can result in strong bipartite entanglement between cavity photons and magnonic modes under two conditions, i.e., $\delta_c \delta_{m}=2g^2$ and $\delta_c=-\delta_{m}$. For the later one, we also show the possibility for producing the bipartite entanglement between two magnonic modes and tripartite entanglement among the cavity photons and two magnonic modes. Combining these two conditions, we further derive a third condition, i.e., $\delta_m^2-\phi^2+2g^2=0$, where the tripartite entanglement can be achieved when two magnonic modes have different resonant frequencies. 
\end{abstract}

\maketitle
Yttrium iron garnet (YIG) materials are good candidates for demonstrating interesting phenomena in quantum optics and condensed matter field of magnetism due to its high Curie temperature, high spin density, low dissipation rate and  good tunability~\cite{bhoi2020roadmap,harder2018cavity}. Particularly, the ferromagnetic resonance (FMR) induced collective spin dynamics gives rise to a new research field of magnonics by combining the meso- and nanoscale science. With modern lithography and sensing techniques, a great amount of fascinating phenomena have been reported theoretically and experimentally, including dynamics of skyrmions~\cite{WangChotorlishvili-3097}, magnetic vortices~\cite{graf2018cavity,osada2018brillouin}, and spin pumping effect~\cite{PhysRevLett.106.216601, PhysRevLett.107.046601, heinrich2011spin, ShiomiLustikova-3102} and so on. All these properties would enable further investigation of quantum optical phenomena in hybrid-quantum systems, integrating of magnonic systems with photons~\cite{ZhangZou-612,harder2018level}, qubits~\cite{tabuchi2015coherent,lachance2020entanglement,cai2019observation}, optomechanics~\cite{LiZhu203601,potts2020magnon,ZhangZou-648} and others.

Besides, the interaction between an ensemble of spins and the cavity field plays an important role in the development of novel hybrid quantum system. In the field of magnonics, photons confined to a cavity mode interact more strongly with a matter polarization, producing the cavity magnon polariton as a new type of quasi particle~\cite{cao2015exchange,kruglyak2010magnonics}. This is because magnon polariton have spin density many orders of magnitude higher than ensembles consisting of atoms, molecules, nitrogen vacancy centers, ion doped crystals and so on. Notably, strong coupling between a Kittel mode in YIG sphere and the photonic mode has been observed at the room temperature~\cite{ZhangWang-460,PhysRevB.97.184420,ZhangZou-324,zollitsch2015high}. 

Recently, this sub field of cavity electromagnonics involving the interaction between magnon modes and the cavity light mode has developed rapidly. Many emergent phenomena have been found, such as cavity spintronics~\cite{BaiHarder-614,BaiHarder-619,yuan2022quantum}, bistability~\cite{WangZhang-1124,zhang2019theory,chai2019thermal}, magnon dark modes~\cite{ZhangZou-324,xiao2019magnon}, magnetically controllable slow light~\cite{KongWang-463,zhao2021phase}, and magnon-induced transparency~\cite{WangLiu-1171,ullah2020tunable}. Particularly, the preparation of entangled states in ferromagnetic materials, e.g., YIG spheres, has attracted great attention. Several methods have been proposed theoretically to realize bipartite and tripartite entanglements~\cite{PhysRevResearch.1.023021, PhysRevLett.124.213604, NairAgarwal-2024, PhysRevA.105.063704,LiZhu203601}. Other applications have also been reported such as the generation of squeezed states of magnons and phonons in cavity magnomechanics~\cite{PhysRevA.99.021801}, and the implementation of nonreciprocal transmission for a microwave field~\cite{PhysRevB.92.104501}.

In this paper, we present a novel method to produce strong bipartite and tripartite entanglements in a two YIG spheres cavity QED system via the second order nonlinearity of microwave field, which can be implemented by utilizing nonlinear materials~\cite{trojanek2010nonlinear,schriever2015second,borghesani2013generation,borghesani2016microwave}, photonic waveguide systems~\cite{nitiss2022optically,guo2016second} as well as the dynamical Casimir effect demonstrated in optomechanics system~\cite{law1994effective,dodonov2020fifty,tanaka2020dynamical,macri2018nonperturbative}. We first consider a special case where two YIG spheres have the same resonant frequencies of magnon modes. We obtain two conditions for realizing strong photon-magnon entanglement, and magnon-magnon entanglement. Moreover, the tripartite entanglement among the cavity field and two magnon modes can also be achieved under one of these two conditions. Then, we consider the case where two magnon modes have different resonant frequencies. We show that a new condition for implementation of tripartite entanglement can be easily derived if we combine these two conditions. In contrast to previous proposals, the second order nonlinearity results in a strong gain of cavity photon numbers, and further enhances the interaction strength between photons and magnons, yielding strong entanglements under weak nonlinearity.

\begin{figure}[htbp]
	\centering
	\includegraphics[width=\linewidth]{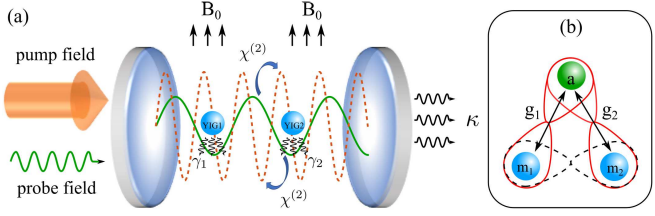}
	\caption{ (a) Schematic of hybrid magnon-cavity QED system. Two YIG spheres with resonant frequencies $\omega_{m1}$ and $\omega_{m2}$ are located inside a microwave cavity driven by a pump field and a auxiliary probe field $\varepsilon_p$. The pump field photons are transferred to the probe field photons via two photon process with nonlinear interaction strength $\Omega$. Here, $\gamma_{1}$ and $\gamma_{2}$ denote the decay rates of two magnonic modes, while  $\kappa$ denotes the decay rate of the cavity mode. (b) Interactions among the subsystems. Two magnon modes linearly couple to the cavity mode with coupling strengths $g_1$ and $g_2$, respectively. The second-order nonlinear interaction result in bipartite entanglements between two magnon modes and the cavity mode, respectively. With some specific conditions, two magnon modes can be entangled, which further leads to tripartite entanglements.
	\label{Fig1:model}
}
\end{figure}
{\it Model.} - As shown in Fig.~\ref{Fig1:model}, we consider a magnon-photon hybrid system where two YIG spheres are placed in a single mode  microwave cavity with resonant frequency $\omega_c$. With current experimental techniques, strong couplings between cavity photons and collective spin excitations in YIG spheres can be achieved~\cite{ZhangWang-460,HueblZollitsch-609,TabuchiIshino-611,GoryachevFarr-545,ZhangZou-612,BaiHarder-614}. In our system, we only take into account the Kittel modes which have spatially uniform profile and subject to giant magnetic moments, i.e., $\mathbf{M}_j=\gamma_e\mathbf{S}^{(j)}/V$. Here,  $\gamma_e=e/m_ec$ is the gyromagnetic ratio for electron spin and $\mathbf{S}^{(j)}\ (j=1,2)$ denotes the collective spin operator of the $j-$th YIG sphere, which couples the external magnetic field and the magnetic field inside the cavity. Thus, the frequency of the Kittel mode in $j-$th YIG sphere $\omega_{mj}=\gamma H_z^{(j)}$, which can be flexibly tuned by adjusting the external magnetic field. By means of the Holstein-Primakoff transform~\cite{PrimakoffHolstein-1184}, the collective spin operators can be approximately represented by the boson creation and annihilation
operators ($\hat{m}_j^\dag$ and $\hat{m}_j$) with $[\hat{m}_j,\hat{m}_j^\dag]=1$. Then, the raising and lowering operators of the spin can be approximately expressed as {$\hat{m}_j\approx \hat{S}_+^{(j)}/\sqrt{(2S)}$ and $\hat{m}_j^\dag\approx \hat{S}_-^{(j)}/\sqrt{(2S)}$} with $\hat{S}_\pm^{(j)}=\sum_{j=1}^N\hat{\sigma}_\pm^{(j,N)}$ and $S=Ns$ being the total spin number of the corresponding collective spin operator, with the total number of spins $N=\rho V$ and the spin number $s=5/2$. Here, we consider a typical yttrium iron garnet with high spin density $\rho=4.22\times10^{27}$ m$^{-3}$ and diameter $d=1$ mm~\cite{PhysRevB.94.224410,WangZhang-1124}. The cavity is driven by a weak auxiliary field with resonant frequency $\omega_p$ and a pump field  with resonant frequency $\omega_P=2\omega_p$. We must point out that the probe field is just used to obtain the conditions for implementation of entanglements, which is not essential in experiments. Under the rotating wave approximation in the frame of the probe field, the Hamiltonian of this magnon cavity system shown in Fig.~\ref{Fig1:model} is (setting $\hbar=1$)
\begin{eqnarray}
	\hat{H}&=&\delta_{c}\hat{a}^{\dag}\hat{a}+\sum_{j=1,2}\left[{\delta_{mj}\hat{m}_{j}^{\dag}\hat{m}_{j}}+g_{j}(\hat{a}\hat{m}_{j}^{\dag}+\hat{a}^{\dag}\hat{m}_{j})\right]\nonumber\\ &&+\Omega(\hat{a}^{2}+{\hat{a}^{\dag 2}})+\varepsilon_p(\hat{a}+\hat{a}^{\dag})
\end{eqnarray}
where $\hat{a}(\hat{a}^{\dag})$ denotes the annihilation (creation) operator of cavity mode, $\delta_{c}=\omega_{c}-\omega_{p}$ and $\delta_{mj}=\omega_{mj}-\omega_{p}$.  $g_j=\sqrt{5N}\gamma_eB_{\rm vac}$ denotes the magnon-cavity coupling with the magnetic field of vacuum $B_{\rm vac}=\sqrt{2\pi\hbar\omega_c/V_{\rm vac}}$. $\varepsilon_p$ is the driving strength of the probe field, and the cavity photons interacts with the pump field via two photon process with nonlinear interaction strength $\Omega$. Such kind of second order nonlinear interaction can be implemented in various quantum systems~\cite{trojanek2010nonlinear,schriever2015second,borghesani2013generation,borghesani2016microwave,nitiss2022optically,guo2016second,law1994effective,dodonov2020fifty,tanaka2020dynamical,macri2018nonperturbative}.

The dynamics of this coupled system is described by the quantum master equation, which reads 
\begin{eqnarray}
	\frac{d\hat{\rho}}{dt}&=&-i[\hat{H}, \hat{\rho}]
	+\frac{\kappa}{2} {\cal \hat{L}}_{\kappa} [\hat{\rho}]+\sum_{j=1,2}\frac{\gamma_j}{2}{\cal \hat{L}}_{\gamma}^{(j)}  [\hat{\rho}],
\end{eqnarray}
where ${\hat{\rho}}$ is the density matrix of the system. The decay terms are given by 
${\cal \hat{L}_{\kappa}} [\hat{\rho}]={\text 2} \hat{a} \hat{\rho} \hat{a}^{\dag}-\hat{a}^{\dag} \hat{a} \hat{\rho} -\hat{\rho} \hat{a}^\dag \hat{a} $ and
${\cal \hat{L}}_{\gamma}^{(j)} [\hat{\rho}]=2\hat{m}_j\hat{\rho} \hat{m}_j^{\dag}-\hat{m}_j^{\dag} \hat{m}_j \hat{\rho} -\hat{\rho} \hat{m}_j^\dag \hat{m}_j $ with the cavity decay rate $\kappa$ and magnon decay rate $\gamma_j$, respectively. 

Then, the time evolution of the bosonic operators, including the thermal fluctuation over and above the mean values, can be described by the quantum Langevin equations (QLEs), which reads
\begin{eqnarray}
\frac{d \hat{a}}{d t}&=&-i(\delta_c-i\kappa) \hat{a}-ig_1\hat{m}_1-ig_2\hat{m}_2-i\varepsilon_p\nonumber\\
& &-2i\Omega \hat{a}^{\dag}+\sqrt{2\kappa} \hat{a}^{in}\\
\frac{d  \hat{m}_1}{d t}&=&-i(\delta_{m1}-i\gamma_1)\hat{m}_1-ig_1 \hat{a}+\sqrt{2\gamma_1} \hat{m}_{1}^{in}\\
\frac{d  \hat{m}_2}{d t}&=&-i(\delta_{m2}-i\gamma_2)\hat{m}_2-ig_2 \hat{a}+\sqrt{2\gamma_2} \hat{m}_{2}^{in}
\end{eqnarray}
where $\hat{a}^{in}$ and $\hat{m}_{j}^{in}$ $(j=1,2)$ denote input quantum noises of the cavity mode and the $j-$th magnon mode, respectively. They obey the following correlations~\cite{GardinerZoller-2844}:
$\langle{\hat{a}^{in}(t)\hat{a}^{in\dag}(t^{'})}\rangle=\delta(t-t^{'})$,
$\langle{\hat{a}^{in\dag}(t)\hat{a}^{in}(t^{'})}\rangle=0$,
$\langle{\hat{m}_{j}^{in}(t)\hat{m}_{j}^{in\dag}(t^{'})}\rangle=\delta(t-t^{'})$,
$\langle{\hat{m}_{j}^{in\dag}(t)\hat{m}_{j}^{in}(t^{'})}\rangle=0$. 
In the following, we set $g_1=g_2\equiv g$, $\gamma_1=\gamma_2\equiv\gamma$ for mathematical simplicity.
Generally, equations (3)-(5) can be solved by using the mean field approximation, i.e., setting an arbitrary operator $\hat{o}=o+\delta \hat{o}\ (o=a, m_1, m_2)$.  Here, $o\equiv\langle \hat{o}\rangle={\rm Tr}(\hat{\rho}\hat{o})$ denotes the average value of the operator $\hat{o}$, while $\delta \hat{o}$ represents the quantum fluctuation above the average value. 

To show the physical mechanism of the entanglement more clearly, we first set $\omega_{m1}=\omega_{m2}\equiv\omega_m$. Under the steady-state approximation, Eqs.~(3)-(5) can be linearized, yielding
\begin{eqnarray}
	&&(\delta_c-i\kappa) a+g(m_1+m_2)+2\Omega a^\ast=-\varepsilon_p,\\
	&&(\delta_{m}-i\gamma)m_j+ga=0,
\end{eqnarray}
where $j=1-2$ and $\delta_m=\omega_m-\omega_p$. The solutions of the above equations are given by 
\begin{subequations}~\label{eq:a}
\begin{eqnarray}
&& a=\frac{\varepsilon_p}{4\Omega^2-|D_0|^2}\left(D_0^\ast-2\Omega\right),\\
&& m_1=m_2=-ga/\delta_{m}, 
\end{eqnarray}
\end{subequations}
where $D_0=\Delta_{c}-2g^2/\Delta_m$ with complex detuning $\Delta_c=\delta_c-i\kappa$ and $\Delta_m=\delta_m-i\gamma$. Then, one can easily obtain the average photon number $n_c\equiv\langle a^\dag a\rangle\approx |a|^2$ and the magnon excitation numbers of the $j-$th YIG sphere $n_{mj}\equiv\langle m_j^\dag m_j\rangle\approx |m_j|^2$. In view of  Eq.~(\ref{eq:a}a) and dropping all decay terms, it is noted that the cavity photons can be excited with its maximal efficiency if the condition 
\begin{equation}
	\delta_c \delta_{m}=2g^2,
\end{equation} 
is satisfied. Simultaneously, the magnon excitation number will also reaches its maximum.

\begin{figure}[htbp]
	\centering
	\includegraphics[width=\linewidth]{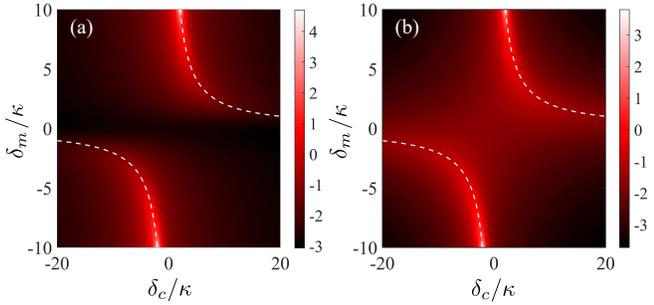}
	\caption{Average photon number $n_c$ [panel (a)] and magnon excitation number $n_{m1}(n_{m2})$ [panel (b)] on a logarithmic scale as functions of normalized detunings $\delta_c/\kappa$ and $\delta_{m}/\kappa$. Here, the white dash curves indicate the condition $\delta_c \delta_{m} = 2g^2$ where maximum value of average photon number and magnon excitation number can be observed. System parameters are given in the text.}
    \label{fig:fig2}
\end{figure}
In Fig.~\ref{fig:fig2}, we show the average photon number $n_c$ [panel (a)] and the magnon excitation number $n_{m1}\ (n_{m2})$ in the first (second) YIG sphere [panel (b)] on a logarithmic scale as functions of the detunings $\delta_c$ and $\delta_{m}$, respectively. Here, we choose $\varepsilon_p=\kappa$ and the nonlinear interaction strength $\Omega/\kappa=0.6$, $\gamma=\kappa$, $g/\kappa=3.2$~\cite{ZhangZou-648}. In Fig.~\ref{fig:fig2}(a), it is clear to see that there exists two excitation branches in the cavity excitation spectrum with the condition $\delta_c\delta_{m}=2g^2$. As shown in panel (b), similar characteristics can be observed in the magnon excitation spectrum. We must point out that the maximal value of the magnon excitation numbers is just about $10^3$ so that weak excitation assumption $m_1,m_2\ll 2Ns$ is satisfied and the H-P approximation is valid. In the following, we will show that how such a weak magnon excitation can lead to strong entanglement between cavity photons and magnons in both YIG spheres.

{\it Bipartite entanglement.} - First, let's consider the bipartite entanglement between cavity photons and magnons by studying the properties of the quadrature fluctuations of the cavity field and the magnon modes, which are defined as 
$\delta X=(\delta \hat{a}+\delta \hat{a}^{\dag})/\sqrt2$, 
$\delta Y=i(\delta \hat{a}^{\dag}-\delta \hat{a})/\sqrt2$, 
$\delta x_1=(\delta \hat{m}_1+\delta \hat{m}_1^{\dag})/\sqrt2$, 
$\delta y_1=i(\delta \hat{m}_1^{\dag}-\delta \hat{m}_1)/\sqrt2$, 
$\delta x_2=(\delta \hat{m}_2+\delta \hat{m}_2^{\dag})/\sqrt2$, and 
$\delta y_2=i(\delta \hat{m}_2^{\dag}-\delta \hat{m}_2)/\sqrt2$. Neglecting higher-order fluctuations of the operators, the evolution of quadrature fluctuations can be described by the linearized QLEs, which reads
\begin{equation}
\dot{f}(t)=Af(t)+\eta
\end{equation}
\allowbreak where $f(t)=[\delta X(t),\delta Y(t),\delta x_1(t),\delta y_1(t),\delta x_2(t),\delta y_2(t)]^{\rm T}$, and $\eta(t)=\left[\sqrt{2\kappa}X^{in}, \sqrt{2\kappa}Y^{in}, \sqrt{2\gamma}x_1^{in}, \sqrt{2\gamma}y_1^{in},\right.$ $\left.\sqrt{2\gamma}x_2^{in}, \sqrt{2\gamma}y_2^{in}\right]^{\rm T}$ is a vector denoting the input noises. The drift matrix is defined as
\begin{equation}
A=\left(\begin{array}{cccccc}
	{-\kappa}&{\delta_{c}-2\Omega}&{0}&{g}&{0}&{g}\\
	{-\delta_{c}-2\Omega}&{-\kappa}&{-g}&{0}&{-g}&{0}\\
	{0}&{g}&{-\gamma}&{\delta_{m}}&{0}&{0}\\
	{-g}&{0}&{-\delta_{m}}&{-\gamma}&{0}&{0}\\
	{0}&{g}&{0}&{0}&{-\gamma}&{\delta_{m}}\\ {-g}&{0}&{0}&{0}&{-\delta_{m}}&{-\gamma}\end{array}\right)
\end{equation}

For such as a system, a $6\times6$ covariance matrix (CM) $\cal {V}$ can be used to describe a continuous variable three-mode Gaussian state. The corresponding element of this CM is defied as $\cal {V}$$_{ij}=\langle f_i(t)f_j(t^{'})+f_j(t^{'})f_i(t)\rangle/2$ $(i,j=1,2,...,6)$. Generally, we can solve the Lyapunov equation to obtain the steady state CM $\cal {V}$~\cite{VitaliTombesi-500,ParksHahn-299}, i.e., 
\begin{equation}
	A{\cal {V}}+{\cal {V}} A^{T}=-D,
\end{equation}
where the diffusion matrix is defined as  
$D=[\kappa$,
$\kappa$,
$\gamma$,
$\gamma$,
$\gamma$,
$\gamma]^{\rm T}$ 
with $D_{ij}\delta(t-t')=\langle \eta_i(t)\eta_j(t')+\eta_j(t')\eta_i(t)\rangle/2$. Then, we calculate the logarithmic negativity~\cite{VidalWerner-490,Plenio-591} to quantitatively measure the bipartite entanglement $E_{\alpha\beta}\ (\alpha,\beta=a,m_1,m_2)$ between any two different modes, i.e.,
\begin{equation}
E_{\alpha\beta}\equiv {\rm max}\{0,-{\rm ln}2\tilde{\nu}_-\}
\end{equation}
where $\tilde{\nu}_{-}={\rm min}\{{\rm eig}(i\Omega_2\tilde{\cal V}_{4})\}$ with $\tilde{\cal V}_{4}=P_{1\mid2}{\cal V}_4P_{1\mid2}$. Here, $\Omega_2=\oplus^2_{j=1}i\sigma_y$, $P_{1\mid2}=\sigma_z\oplus I$ and ${\cal V}_4$ is a $4\times 4$ CM of arbitrary two subsystems in this three-mode system, which can be obtained by  deleting rows and columns of irrelevant modes in CM ${\cal V}$. $\sigma_y$ and $\sigma_z$ are the Pauli matrices. As usual, $E_{\alpha\beta}>0$ denotes the existence of bipartite entanglement. 

\begin{figure}[htbp]
	\centering
	\includegraphics[width=\linewidth]{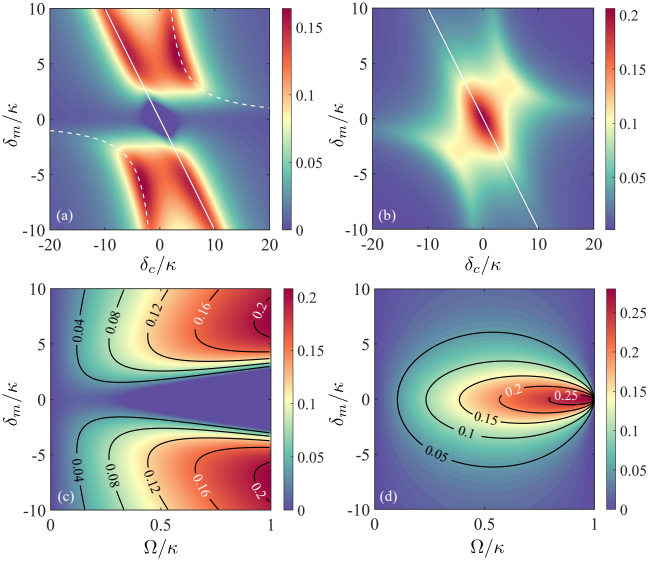}
	\caption{Density plot of bipartite entanglement $E_{am1}=E_{am2}$ [panel (a)] and $E_{m1m2}$ [panel (b)] versus normalized detunings $\delta_c/\kappa$ and $\delta_{m}/\kappa$. White dash curves indicate the condition $\delta_c \delta_{m}=2g^2$, while white solid lines indicate the condition $\delta_c =-\delta_{m}$. Panels (c) and (d) show the density plot of bipartite entanglements $E_{am1}=E_{am2}$ and $E_{m1m2}$ against the normalized nonlinear interaction strength $\Omega/\kappa$ and detuning $\delta_{m}/\kappa$ by fixing $\delta_c=-\delta_{m}$.}
	\label{fig:fig3}
\end{figure}
Fig.~\ref{fig:fig3}(a) shows the bipartite entanglement $E_{\rm am1}$ ($E_{\rm am2}$) between the magnon mode in the first (second) YIG sphere and the cavity mode as functions of the detunings $\delta_c$ and $\delta_{m}$, respectively. The system parameters are the same as those used in Fig.~\ref{fig:fig2}. Obviously, strong bipartite entanglements between the magnon mode and the cavity mode occur under two different conditions. One is $\delta_c\delta_{m}=2g^2$ (white dashed curves) as demonstrated in Fig.~\ref{fig:fig2}. The other (white solid line) is 
\begin{equation}
	\delta_c=-\delta_{m},
\end{equation} 
which can be understood by exploring the system in bare state picture. Considering two bare states labeled by $|N_c,N_{mj}\rangle$ and $|N_c-1,N_{mj}+1\rangle$, a bipartite entanglement state such as  $(|N_c,N_{m1}\rangle+|N_c-1,N_{m1}+1\rangle)/\sqrt{2}$ will be produced if both states have the same excitation probabilities. Thus, the probe field frequency must satisfy  $\omega_p=(\omega_c+\omega_{mj})/2$, yielding $\delta_c=-\delta_{m}$. It is also noted that, in a small regime near $\delta_c=\delta_{m}=0$ (i.e., $\omega_c=\omega_{m1}=\omega_{m2}$), the photon mode and magnon mode are not entangled and the bipartite entanglement $E_{am1}=E_{am2}=0$ since these two states can not be distinguished [see panel (a)]. In Fig.~\ref{fig:fig3}(b), we show the bipartite entanglement $E_{\rm m1m2}$ between two magnonic modes. In contrast to the bipartite entanglement $E_{amj}$, the bipartite entanglement $E_{m1m2}$ only appears in the regime near the condition of $\delta_c=-\delta_{m1}$. In particular, the maximal bipartite entanglement between two magnonic modes occurs at center point with $\omega_c=\omega_{m1}=\omega_{m2}$. However, at this point, the magnon mode and cavity mode are not entangled, which is coincided with the system driven by a JPA process~\cite{NairAgarwal-2024}. 

Here, we must point out that the second order nonlinearity is the key for generating bipartite entanglements. To show this point, we fix $\delta_c=-\delta_{m}$. Fig.~\ref{fig:fig3}(c) and Fig.~\ref{fig:fig3}(d) show the $E_{am1}$ ($E_{am2}$) and $E_{m1m2}$ against the normalized detuning $\delta_{m}/\kappa$ and nonlinear interaction strength $\Omega/\kappa$, respectively. Obviously, $E_{am1}=E_{am2}=E_{m1m2}=0$ (non-entanglement) if the nonlinear interaction strength $\Omega=0$. Bipartite entanglements $E_{am1}$ ($E_{am2}$) and $E_{m1m2}$ are significantly enhanced as the nonlinear interaction strength $\Omega$ increases. It is noted that this second order nonlinearity induced bipartite entanglements reach up to $0.1$ even for a weak nonlinear interaction strength, e.g., $\Omega/\kappa=0.5$. It is as strong as the phonon induced bipartite entanglements reported in Ref.~\cite{LiZhu-339}, where the average magnon excitation number is above $10^7$ to acquire strong nonlinear effect. Compared with panels (c) and (d), we notice that it is possible to find some regimes where mutual bipartite entanglements (i.e., tripartite entanglement with non-zero $E_{am1}$, $E_{am2}$ and $E_{m_1m_2}$) can be achieved when the driving field is detuned.

{\it Tripartite entanglement.} - To verify this feature, we adopt the minimum residual contangle as a $bona\ fide$ quantification of tripartite entanglement~\cite{AdessoIlluminati-1195,AdessoIlluminati-1196}. Here, contangle is a CV analogue of tangle for discrete-variable tripartite entanglement, and the minimum residual contangle is given by
\begin{equation}
R_{\tau}^{\rm min}\equiv \min\{R_{\tau}^{a\mid m_1m_2},R_{\tau}^{m_1\mid am_2},R_{\tau}^{m_2\mid am_1}\}
\end{equation}
where $R_{\tau}^{i\mid jk}\equiv C_{i\mid jk}-C_{i\mid j}-C_{i\mid k}\geq 0$ $(i,j,k=a,m_1,m_2)$ denotes the residual contangle with $C_{u\mid v}$ being the contangle of subsystems $u$ and $v$ ($v$ can contain one or two modes). Here, we consider that $v$ contains two modes, and the contangle $C_{i\mid jk}=[{\rm max}\{0,-{\rm ln}(2\tilde{\nu}_-)\}]^2$, where $\tilde{\nu}_-\equiv {\rm min}\{{\rm eig}(i\Omega_3\tilde{\cal V}_{6})\}$ with $\Omega_3=\oplus^3_{j=1}i\sigma_y$ and $\tilde{\cal {V}}_{6}=P_{i\mid jk}{\cal {V}} P_{i\mid jk}$. Here, $P_{1\mid 23}=\sigma_z\oplus I\oplus I$, $P_{2\mid 13}=I\oplus \sigma_z\oplus I$ and $P_{3\mid 12}=I\oplus I \oplus \sigma_z$  denote partial transposition matrices. Thus,  $R_{\tau}^{min}>0$ represents the existence of genuine tripartite entanglement in the system.
\begin{figure}[htbp]
	\centering
	\includegraphics[width=\linewidth]{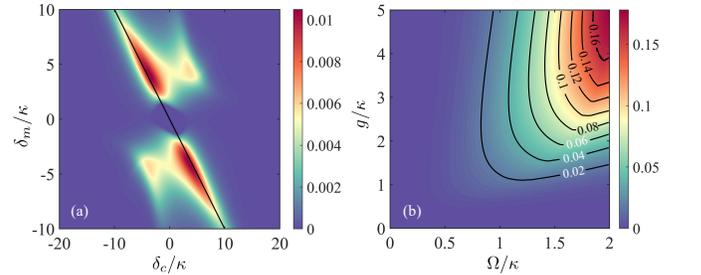}
	\caption{(a) The tripartite entanglement $R_{\tau}^{\rm min}$ verses normalized detunings $\delta_c/\kappa$ and $\delta_{m}/\kappa$. Black solid line indicates the condition $\delta_c =-\delta_{m}$. (b) The maximal Tripartite entanglement $R_{\tau}^{\rm min}$ is plotted as functions of the normalized nonlinear interaction strength $\Omega/\kappa$ and coupling strength $g/\kappa$ by fixing $\delta_c =-\delta_{m}$ and scanning magnon detuning $\delta_{m}$ over a wide range.}
	\label{fig:fig4}
\end{figure}

Next, we will discuss the possibility for generating a strong tripartite entanglement via the second-order nonlinearity. In Fig.~\ref{fig:fig4}(a), we show the tripartite entanglement $R^{\rm min}_\tau$ versus detunings $\delta_c$ and $\delta_{m}$, respectively. Here, the system parameters are the same as those used in Fig.~\ref{fig:fig2}, and the black solid line indicates the condition $\delta_c=-\delta_{m}$. As expected, strong tripartite entanglements occur near this condition with a non-zero cavity/magnon detuning. Fig.~\ref{fig:fig4}(b) shows more clearly the presence of tripartite entanglement by setting $\delta_c=-\delta_{m}$. Here, we plot the maximal tripartite entanglement against the normalized coupling strength $g/\kappa$ and the nonlinear interaction strength $\Omega/\kappa$ by scanning the magnon detuning $\delta_{m}$ over a wide range. Obviously, strong tripartite entanglement can be produced when a set of suitable photon-magnon interaction strength $g$ and nonlinear interaction strength $\Omega$ is chosen. It is noted that the tripartite entanglement is stronger than the OPA induced entanglement reported in Ref.~\cite{PhysRevA.105.063704}.

\begin{figure}[htbp]
	\centering
    \includegraphics[width=\linewidth]{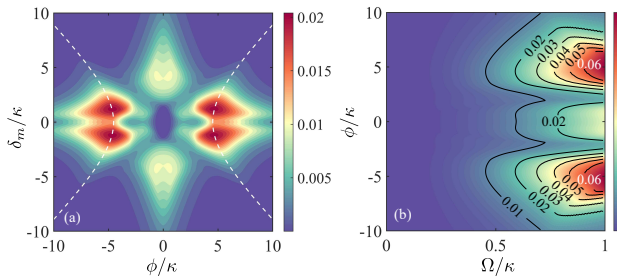}
	\caption{(a) The tripartite entanglement $R_{\tau}^{\rm min}$ verses the frequency difference between two magnonic modes $\phi/\kappa$ and the detuning $\delta_m/\kappa$. Here, we choose $\delta_c=-\delta_m$ and white dash curves indicate the condition $\delta_m^2-\phi^2+2g^2=0$. (b) The maximal tripartite entanglement $R_{\tau}^{\rm min}$ verses the nonlinear interaction strength $\Omega/\kappa$ and the frequency difference between two magnon modes $\phi/\kappa$ by setting $\delta_m^2-\phi^2+2g^2=0$ and scanning the detuning $\delta_{m}$ over a wide range.}
	\label{fig:fig5}
\end{figure}

Finally, let's study the presence of tripartite entanglements when two magnon modes have different resonant frequencies. To show the properties of tripartite features more clearly, we define the average magnon frequency  $\bar{\omega}_m\equiv(\omega_{m1}+\omega_{m2})/2$ and the frequency difference  $\phi\equiv(\omega_{m1}-\omega_{m2})/2$. Then, the detunings are given by $\delta_{m1}=\delta_m+\phi$ and $\delta_{m2}=\delta_m-\phi$ with $\delta_m=\bar{\omega}_m-\omega_p$. Based on the above analysis, it is found that there exist two different conditions to realize strong bipartite entanglement between photon mode and single magnon mode (see Eqs~(\ref{eq:a})). Therefore, the presence of tripartite entanglement with two different magnon modes can be predicted if the frequency of the first magnon mode satisfies $\delta_c=-\delta_{m1}$, while the frequency of the second magnon mode satisfies $\delta_c\delta_{m2}=2g^2$ simultaneously. Combining these two conditions, one can easily obtain 
\begin{equation}
	\delta_m^2-\phi^2+2g^2=0
\end{equation}
for achieving strong tripartite entanglement. To verify this prediction, we plot $R_{\tau}^{\rm min}$ versus the detunings $\delta_m$ and $\phi$ in Fig.~\ref{fig:fig5}(a). Here, we set $\delta_c=-\delta_m$ and other system parameters are the same as those used in Fig.~\ref{fig:fig2}. It is clear to see that the maximal tripartite entanglements appear at the condition $\delta_m^2-\phi^2+2g^2=0$ indicated by white dashed curves. In Fig.~\ref{fig:fig5}(b), we show the influence of the nonlinear interaction strength $\Omega$ on the presence and quality of the tripartite entanglement. Here, we plot the optimal tripartite entanglement versus $\phi/\kappa$ and $\Omega/\kappa$ by scanning the detuning $\delta_m$ over a wide range. It is found that the tripartite entanglement with two different magnon modes can also be realized under weak nonlinear interaction and it can be significantly enhanced as the nonlinear interaction strength $\Omega$ increases.


In conclusion, we have proposed a scheme to generate bipartite and tripartite entanglement in a hybrid magnon-cavity QED system, where the cavity photons are generated via the two photon process. In the presence of the second order nonlinearity, we show that the strong bipartite entanglement between the cavity mode and magnon mode can be achieved under two conditions. One is $\delta_c \delta_{m} = 2g^2$, the other is $\delta_c =-\delta_{m}$. For the second condition, it is also possible to produce mutual entanglement between two magnon modes near the resonance region. Besides, we show that the optimal tripartite entanglement can be implemented if the second condition $\delta_c =-\delta_{m}$ is fulfilled. Combining these two conditions, we derive the third condition for realize the optimal tripartite entanglement associated with two magnon modes with different resonant frequencies, i.e., $\delta_m^2-\phi^2+2g^2=0$. All these conditions are helpful for experimentists to realize macroscopic bipartite and tripartite entanglements in hybrid magnon cavity QED systems.

\begin{acknowledgments}
We thank Dr. Jie Li at Zhejiang University for helpful discussion. This work has been supported by the National Natural Science Foundation of China (Grant No. 61975154).
\end{acknowledgments}


\bibliographystyle{unsrt}

\bibliography{Ref}

\begin{thebibliography}{10}

\bibitem{bhoi2020roadmap}
B.~Bhoi and S.~K. Kim.
\newblock Roadmap for photon-magnon coupling and its applications.
\newblock In {\em Solid State Phys.}, volume~71, pages 39--71. Elsevier, 2020.

\bibitem{harder2018cavity}
M.~Harder and C.~M. Hu.
\newblock Cavity spintronics: an early review of recent progress in the study
  of magnon--photon level repulsion.
\newblock {\em Solid State Phys.}, 69:47--121, 2018.

\bibitem{WangChotorlishvili-3097}
X.~Wang, L.~Chotorlishvili, V.~K. Dugaev, A.~Ernst, I.~V. Maznichenko,
  N.~Arnold, C.~Jia, J.~Berakdar, I.~Mertig, and J.~Barnaś.
\newblock The optical tweezer of skyrmions.
\newblock {\em npj Comput. Mater.}, 6(1):140, 2020.

\bibitem{graf2018cavity}
J.~Graf, H.~Pfeifer, F.~Marquardt, and S.~V. Kusminskiy.
\newblock Cavity optomagnonics with magnetic textures: Coupling a magnetic
  vortex to light.
\newblock {\em Phys. Rev. B}, 98(24):241406, 2018.

\bibitem{osada2018brillouin}
A~Osada, A~Gloppe, R~Hisatomi, A~Noguchi, R~Yamazaki, M~Nomura, Y~Nakamura, and
  K~Usami.
\newblock Brillouin light scattering by magnetic quasivortices in cavity
  optomagnonics.
\newblock {\em Phys. Rev. Lett.}, 120(13):133602, 2018.

\bibitem{PhysRevLett.106.216601}
C.~W. Sandweg, Y.~Kajiwara, A.~V. Chumak, A.~A. Serga, V.~I. Vasyuchka, M.~B.
  Jungfleisch, E.~Saitoh, and B.~Hillebrands.
\newblock Spin pumping by parametrically excited exchange magnons.
\newblock {\em Phys. Rev. Lett.}, 106:216601, 2011.

\bibitem{PhysRevLett.107.046601}
F.~D. Czeschka, L.~Dreher, M.~S. Brandt, M.~Weiler, M.~Althammer, I.-M. Imort,
  G.~Reiss, A.~Thomas, W.~Schoch, W.~Limmer, H.~Huebl, R.~Gross, and S.~T.~B.
  Goennenwein.
\newblock Scaling behavior of the spin pumping effect in ferromagnet-platinum
  bilayers.
\newblock {\em Phys. Rev. Lett.}, 107:046601, 2011.

\bibitem{heinrich2011spin}
B.~Heinrich, C.~Burrowes, E.~Montoya, B.~Kardasz, E.~Girt, Y.~Y. Song, Y.~Y.
  Sun, and M.~Z. Wu.
\newblock Spin pumping at the magnetic insulator (yig)/normal metal (au)
  interfaces.
\newblock {\em Phys. Rev. Lett.}, 107(6):066604, 2011.

\bibitem{ShiomiLustikova-3102}
Y.~Shiomi, J.~Lustikova, S.~Watanabe, D.~Hirobe, S.~Takahashi, and E.~Saitoh.
\newblock Spin pumping from nuclear spin waves.
\newblock {\em Nat. Phys.}, 15(1):22--26, 2019.

\bibitem{ZhangZou-612}
X.~F. Zhang, C.~L. Zou, L.~Jiang, and H.~X. Tang.
\newblock Strongly coupled magnons and cavity microwave photons.
\newblock {\em Phys. Rev. Lett.}, 113(15):156401, 2014.

\bibitem{harder2018level}
M.~Harder, Y.~Yang, B.~M. Yao, C.~H. Yu, J.~W. Rao, Y.~S. Gui, R.~L. Stamps,
  and C.~M. Hu.
\newblock Level attraction due to dissipative magnon-photon coupling.
\newblock {\em Phys. Rev. Lett.}, 121(13):137203, 2018.

\bibitem{tabuchi2015coherent}
Y.~Tabuchi, S.~Ishino, A.~Noguchi, T.~Ishikawa, R.~Yamazaki, K.~Usami, and
  Y.~Nakamura.
\newblock Coherent coupling between a ferromagnetic magnon and a
  superconducting qubit.
\newblock {\em Science}, 349(6246):405--408, 2015.

\bibitem{lachance2020entanglement}
D.~Lachance-Quirion, S.~P. Wolski, Y.~Tabuchi, S.~Kono, K.~Usami, and
  Y.~Nakamura.
\newblock Entanglement-based single-shot detection of a single magnon with a
  superconducting qubit.
\newblock {\em Science}, 367(6476):425--428, 2020.

\bibitem{cai2019observation}
W.~Z. Cai, J.~X. Han, F.~Mei, Y.~Xu, Y.~W. Ma, X.~G. Li, H.~Y. Wang, Y.~P.
  Song, Z.~Y. Xue, and Z.~Q. Yin.
\newblock Observation of topological magnon insulator states in a
  superconducting circuit.
\newblock {\em Phys. Rev. Lett.}, 123(8):080501, 2019.

\bibitem{LiZhu203601}
J.~Li, S.~Y. Zhu, and G.~S. Agarwal.
\newblock Magnon-photon-phonon entanglement in cavity magnomechanics.
\newblock {\em Phys. Rev. Lett.}, 121:203601, 2018.

\bibitem{potts2020magnon}
C.~A. Potts, Victor A. S.~V. Bittencourt, S.~V. Kusminskiy, and J.~P. Davis.
\newblock Magnon-phonon quantum correlation thermometry.
\newblock {\em Phys. Rev. Appl.}, 13(6):064001, 2020.

\bibitem{ZhangZou-648}
X.~F. Zhang, C.~L. Zou, L.~Jiang, and H.~X. Tang.
\newblock Cavity magnomechanics.
\newblock {\em Sci. Adv.}, 2(3):e1501286, 2016.

\bibitem{cao2015exchange}
Y.~S. Cao, P.~Yan, H.~Huebl, S.~T.~B. Goennenwein, and G.~E.~W. Bauer.
\newblock Exchange magnon-polaritons in microwave cavities.
\newblock {\em Phys. Rev. B}, 91(9):094423, 2015.

\bibitem{kruglyak2010magnonics}
V.~V. Kruglyak, S.~O. Demokritov, and D.~Grundler.
\newblock Magnonics.
\newblock {\em J. Phys. D: Appl. Phys.}, 43(26):264001, 2010.

\bibitem{ZhangWang-460}
D.~K. Zhang, X.~M. Wang, T.~F. Li, X.~Q. Luo, W.~D. Wu, F.~Nori, and J.~Q. You.
\newblock Cavity quantum electrodynamics with ferromagnetic magnons in a small
  yttrium-iron-garnet sphere.
\newblock {\em npj Quan. Inf.}, 1(1):15014, 2015.

\bibitem{PhysRevB.97.184420}
I.~Boventer, M.~Pfirrmann, J.~Krause, Y.~Schön, M.~Kläui, and M.~Weides.
\newblock Complex temperature dependence of coupling and dissipation of cavity
  magnon polaritons from millikelvin to room temperature.
\newblock {\em Phys. Rev. B}, 97:184420, 2018.

\bibitem{ZhangZou-324}
X.~F. Zhang, C.~L. Zou, N.~Zhu, F.~Marquardt, L.~Jiang, and H.~X. Tang.
\newblock Magnon dark modes and gradient memory.
\newblock {\em Nat. Commun.}, 6(1):8914, 2015.

\bibitem{zollitsch2015high}
C.~W. Zollitsch, K.~Mueller, D.~P. Franke, S.~T.~B. Goennenwein, M.~S. Brandt,
  R.~Gross, and H.~Huebl.
\newblock High cooperativity coupling between a phosphorus donor spin ensemble
  and a superconducting microwave resonator.
\newblock {\em Appl. Phys. Lett.}, 107(14):142105, 2015.

\bibitem{BaiHarder-614}
L.~Bai, M.~Harder, Y.~P. Chen, X.~Fan, J.~Q. Xiao, and C.~M. Hu.
\newblock Spin pumping in electrodynamically coupled magnon-photon systems.
\newblock {\em Phys. Rev. Lett.}, 114(22):227201, 2015.

\bibitem{BaiHarder-619}
L.~H. Bai, M.~Harder, P.~Hyde, Z.H. Zhang, C.~M. Hu, Y.~P. Chen, and J.~Q.
  Xiao.
\newblock Cavity mediated manipulation of distant spin currents using a
  cavity-magnon-polariton.
\newblock {\em Phys. Rev. Lett.}, 118(21):217201, 2017.

\bibitem{yuan2022quantum}
H.~Y. Yuan, Y.~S. Cao, A.~Kamra, R.~A. Duine, and P.~Yan.
\newblock Quantum magnonics: when magnon spintronics meets quantum information
  science.
\newblock {\em Phys. Rep.}, 965:1--74, 2022.

\bibitem{WangZhang-1124}
Y.~P. Wang, G.~Q. Zhang, D.~k. Zhang, T.~F. Li, C.~M. Hu, and J.~Q. You.
\newblock Bistability of cavity magnon polaritons.
\newblock {\em Phys. Rev. Lett.}, 120(5):057202, 2018.

\bibitem{zhang2019theory}
G.~Q. Zhang, Y.~P. Wang, and J.~Q. You.
\newblock Theory of the magnon kerr effect in cavity magnonics.
\newblock {\em Sci. China Phys. Mech.}, 62(8):1--11, 2019.

\bibitem{chai2019thermal}
C.~Z. Chai, X.~X. Hu, C.~L. Zou, G.~C. Guo, and C.~H. Dong.
\newblock Thermal bistability of magnon in yttrium iron garnet microspheres.
\newblock {\em Appl. Phys. Lett.}, 114(2):021101, 2019.

\bibitem{xiao2019magnon}
Y.~Xiao, X.~H. Yan, Y.~Zhang, V.~L. Grigoryan, C.~M. Hu, H.~Guo, and K.~Xia.
\newblock Magnon dark mode of an antiferromagnetic insulator in a microwave
  cavity.
\newblock {\em Phys. Rev. B}, 99(9):094407, 2019.

\bibitem{KongWang-463}
C.~Kong, B.~Wang, Z.~X. Liu, H.~Xiong, and Y.~Wu.
\newblock Magnetically controllable slow light based on magnetostrictive
  forces.
\newblock {\em Opt. Express}, 27(4):5544, 2019.

\bibitem{zhao2021phase}
J.~Zhao, L.~H. Wu, T.~F. Li, Y.~X. Liu, F.~Nori, Y.~L. Liu, and J.~F. Du.
\newblock Phase-controlled pathway interferences and switchable fast-slow light
  in a cavity-magnon polariton system.
\newblock {\em Phys. Rev. Appl.}, 15(2):024056, 2021.

\bibitem{WangLiu-1171}
B.~Wang, Z.~X. Liu, C.~Kong, H.~Xiong, and Y.~Wu.
\newblock Magnon-induced transparency and amplification in pt-symmetric
  cavity-magnon system.
\newblock {\em Opt. Express}, 26(16):20248, 2018.

\bibitem{ullah2020tunable}
K.~Ullah, M.~T. Naseem, and {\"O}.~M{\"u}stecapl{\i}o{\u{g}}lu.
\newblock Tunable multiwindow magnomechanically induced transparency, fano
  resonances, and slow-to-fast light conversion.
\newblock {\em Phys. Rev. A}, 102(3):033721, 2020.

\bibitem{PhysRevResearch.1.023021}
Z.~D. Zhang, M.~O. Scully, and G.~S. Agarwal.
\newblock Quantum entanglement between two magnon modes via kerr nonlinearity
  driven far from equilibrium.
\newblock {\em Phys. Rev. Res.}, 1:023021, 2019.

\bibitem{PhysRevLett.124.213604}
M.~Yu, H.~Shen, and J.~Li.
\newblock Magnetostrictively induced stationary entanglement between two
  microwave fields.
\newblock {\em Phys. Rev. Lett.}, 124:213604, 2020.

\bibitem{NairAgarwal-2024}
Jayakrishnan M.~P. Nair and G.~S. Agarwal.
\newblock Deterministic quantum entanglement between macroscopic ferrite
  samples.
\newblock {\em Appl. Phys. Lett.}, 117(8):084001, 2020.

\bibitem{PhysRevA.105.063704}
B.~Hussain, S.~Qamar, and M.~Irfan.
\newblock Entanglement enhancement in cavity magnomechanics by an optical
  parametric amplifier.
\newblock {\em Phys. Rev. A}, 105:063704, 2022.

\bibitem{PhysRevA.99.021801}
J.~Li, S.~Y. Zhu, and G.~S. Agarwal.
\newblock Squeezed states of magnons and phonons in cavity magnomechanics.
\newblock {\em Phys. Rev. A}, 99:021801, 2019.

\bibitem{PhysRevB.92.104501}
A.~L. Pankratov, K.~G. Fedorov, M.~Salerno, S.~V. Shitov, and A.~V. Ustinov.
\newblock Nonreciprocal transmission of microwaves through a long josephson
  junction.
\newblock {\em Phys. Rev. B}, 92:104501, 2015.

\bibitem{trojanek2010nonlinear}
F.~Troj{\'a}nek, K.~{\v{Z}}{\'\i}dek, B.~Dzur{\v{n}}{\'a}k, M.~Koz{\'a}k, and
  P.~Mal{\`y}.
\newblock Nonlinear optical properties of nanocrystalline diamond.
\newblock {\em Opt. Express}, 18(2):1349--1357, 2010.

\bibitem{schriever2015second}
C.~Schriever, F.~Bianco, M.~Cazzanelli, M.~Ghulinyan, C.~Eisenschmidt,
  J.~de~Boor, A.~Schmid, J.~Heitmann, L.~Pavesi, and J.~Schilling.
\newblock Second-order optical nonlinearity in silicon waveguides:
  Inhomogeneous stress and interfaces.
\newblock {\em Adv. Opt. Mater.}, 3(1):129--136, 2015.

\bibitem{borghesani2013generation}
A.~F. Borghesani, C.~Braggio, and G.~Carugno.
\newblock Generation of microwave radiation by nonlinear interaction of a
  high-power, high-repetition rate, 1064 nm laser in ktiopo 4 crystals.
\newblock {\em Opt. Lett.}, 38(21):4465--4468, 2013.

\bibitem{borghesani2016microwave}
A.~F. Borghesani, C.~Braggio, and M.~Guarise.
\newblock Microwave emission by nonlinear crystals irradiated with a
  high-intensity, mode-locked laser.
\newblock {\em J. Opt.}, 18(6):065503, 2016.

\bibitem{nitiss2022optically}
E.~Nitiss, J.~Q. Hu, A.~Stroganov, and C.~S. Br{\`e}s.
\newblock Optically reconfigurable quasi-phase-matching in silicon nitride
  microresonators.
\newblock {\em Nat. Photonics}, 16(2):134--141, 2022.

\bibitem{guo2016second}
X.~Guo, C.~L. Zou, and H.~X. Tang.
\newblock Second-harmonic generation in aluminum nitride microrings with
  2500\%/w conversion efficiency.
\newblock {\em Optica}, 3(10):1126--1131, 2016.

\bibitem{law1994effective}
C.~K. Law.
\newblock Effective hamiltonian for the radiation in a cavity with a moving
  mirror and a time-varying dielectric medium.
\newblock {\em Phys. Rev. A}, 49(1):433, 1994.

\bibitem{dodonov2020fifty}
V.~Dodonov.
\newblock Fifty years of the dynamical casimir effect.
\newblock {\em Physics}, 2(1):67--104, 2020.

\bibitem{tanaka2020dynamical}
S.~Tanaka and K.~Kanki.
\newblock The dynamical casimir effect in a dissipative optomechanical cavity
  interacting with photonic crystal.
\newblock {\em Physics}, 2(1):34--48, 2020.

\bibitem{macri2018nonperturbative}
V.~Macr{\`\i}, A.~Ridolfo, O.~Di~Stefano, A.~F. Kockum, F.~Nori, and
  S.~Savasta.
\newblock Nonperturbative dynamical casimir effect in optomechanical systems:
  vacuum casimir-rabi splittings.
\newblock {\em Phys. Rev. X}, 8(1):011031, 2018.

\bibitem{HueblZollitsch-609}
H.~Huebl, C.~W. Zollitsch, J.~Lotze, F.~Hocke, M.~Greifenstein, A.~Marx,
  R.~Gross, and S.~T.~B. Goennenwein.
\newblock High cooperativity in coupled microwave resonator ferrimagnetic
  insulator hybrids.
\newblock {\em Phys. Rev. Lett.}, 111(12):127003, 2013.

\bibitem{TabuchiIshino-611}
Y.~Tabuchi, S.~Ishino, T.~Ishikawa, R.~Yamazaki, K.~Usami, and Y.~Nakamura.
\newblock Hybridizing ferromagnetic magnons and microwave photons in the
  quantum limit.
\newblock {\em Phys. Rev. Lett.}, 113(8):083603, 2014.

\bibitem{GoryachevFarr-545}
M.~Goryachev, W.~G. Farr, D.~L. Creedon, Y.~Fan, M.~Kostylev, and M.~E. Tobar.
\newblock High-cooperativity cavity qed with magnons at microwave frequencies.
\newblock {\em Phys. Rev. Appl.}, 2(5):054002, 2014.

\bibitem{PrimakoffHolstein-1184}
T.~Holstein and H.~Primakoff.
\newblock Field dependence of the intrinsic domain magnetization of a
  ferromagnet.
\newblock {\em Phys. Rev.}, 58(12):1098, 1940.

\bibitem{PhysRevB.94.224410}
Y.P. Wang, G.~Q. Zhang, D.~K Zhang, X.~Q. Luo, W.~Xiong, S.~P. Wang, T.~F. Li,
  C.~M. Hu, and J.~Q. You.
\newblock Magnon kerr effect in a strongly coupled cavity-magnon system.
\newblock {\em Phys. Rev. B}, 94:224410, 2016.

\bibitem{GardinerZoller-2844}
C.~W. Gardiner and P.~Zoller.
\newblock {\em Quantum Noise}.
\newblock Springer- Verlag Berlin, 2000.

\bibitem{VitaliTombesi-500}
D.~Vitali, P.~Tombesi, M.~J. Woolley, A.~C. Doherty, and G.~J. Milburn.
\newblock Entangling a nanomechanical resonator and a superconducting microwave
  cavity.
\newblock {\em Phys. Rev. A}, 76(4):042336, 2007.

\bibitem{ParksHahn-299}
P.~C. Parks and V.~Hahn.
\newblock {\em Stability theory}.
\newblock Springer New York, 1993.

\bibitem{VidalWerner-490}
G.~Vidal and R.~F. Werner.
\newblock Computable measure of entanglement.
\newblock {\em Phys. Rev. A}, 65(3A):032314, 2002.

\bibitem{Plenio-591}
M.~B. Plenio.
\newblock Logarithmic negativity: a full entanglement monotone that is not
  convex.
\newblock {\em Phys. Rev. Lett.}, 95(9):090503, 2005.

\bibitem{LiZhu-339}
J.~Li and S.~Y. Zhu.
\newblock Entangling two magnon modes via magnetostrictive interaction.
\newblock {\em New J. Phys.}, 21(8):85001, 2019.

\bibitem{AdessoIlluminati-1195}
G.~Adesso and F.~Illuminati.
\newblock Continuous variable tangle, monogamy inequality, and entanglement
  sharing in gaussian states of continuous variable systems.
\newblock {\em New J. Phys.}, 8:15, 2006.

\bibitem{AdessoIlluminati-1196}
G.~Adesso and F.~Illuminati.
\newblock Entanglement in continuous variable systems: recent advances and
  current perspectives.
\newblock {\em J. Phys. A-Math. Theo.}, 40(28):7821, 2007.

\end{thebibliography}

\end{document}